\documentclass[aps,prd,twocolumn]{revtex4-1}

\usepackage[bookmarks=false,colorlinks=true]{hyperref}
\usepackage{amsmath}
\usepackage{hyperref}

\newcommand{\be}{\begin{equation}}
\newcommand{\ee}{\end{equation}}
\newcommand{\sfrac}[2]{{\textstyle\frac{#1}{#2}}}

\begin{document}

\title{On symmetries of the generalized Calogero model and Polychronakos-Frahm chain}
\author{Tigran Hakobyan}
\email{tigran.hakobyan@ysu.am}
\affiliation{Yerevan State University, 1 Alex Manoogian St., Yerevan, 0025, Armenia}

\begin{abstract}
The symmetry of the generalized Polychronakos-Frahm chain is obtained from
the Dunkl-operator deformation of the unitary algebra, which  describes
the symmetry of the generalized Calogero model.
\end{abstract}

\maketitle

\section{Introduction}

Integrable chains with inverse-square interactions were first introduced by Haldane and Shastry for equidistantly localized spins
on a cycle  \cite{haldane88,shastry88}. The Haldane-Shastry model  extends the nearest-neighbor Heisenberg  chain to long-range interactions
preserving the integrability.
It can be considered also as a discrete analog of the integrable Calogero-Sutherland system describing particles
confined on a cycle with a trigonometric inverse-square
interaction \cite{suther}.
The rational version of the latter 
had been  proposed earlier and solved by
Calogero \cite{cal}.  A related discrete system was suggested and studied by Polychronakos and Frahm \cite{poly93,frahm}.
There are various integrable  extensions of Calogero-type systems, in particular, to hyperbolic
and elliptic potentials, particles with internal spin degrees of freedom \cite{min-poly}, finite reflection groups,
Coulomb potential \cite{khare},
etc. (see Refs. \cite{rev-olsh,rev-poly} for a review).

Apart from the Liouville integrals \cite{moser}, the unbound Calogero system (Calogero-Moser model)
possesses additional
integrals of motion ensuring maximal superintegralility \cite{woj83,kuznetsov,gonera}.
This property is retained in the presence of  the oscillator \cite{kosinski}
and Coulomb potentials including the spaces  with constant curvature \cite{CalCoul}.

There is an elegant way of solving and constructing
of integrals  for quantum systems based on the Dunkl exchange operator \cite{dunkl}.
The inverse-square (Calogero)
interaction is hiding in the covariant
derivative  provided by such an operator \cite{poly92,brink}.
The related connection is flat but includes
the two-particle exchanges, making the constructed Hamiltonian,  called sometimes
a \emph{generalized} Hamiltonian,  and related observables essentially nonlocal.
The unwanted exchanges disappear  on the bosonic or fermionic states,
recovering the original Calogero Hamiltonian.
Likewise, the symmetries and wave functions of the  generalized Hamiltonian are  Dunkl-operator
deformations of those for the underlying system without the inverse-square interaction.
Once they are constructed,  the integrals of motion of  the original Calogero Hamiltonian are
recovered as symmetric polynomials in symmetry generators of the generalized system.

In this context, the generalized  Calogero Hamiltonian remains invariant with respect
to the Dunkl-operator deformation of angular momentum generators.
In addition, the hidden symmetry is provided by the  Dunkl analog of
the Fradkin tensor.  Both of them form an extension of the conventional
unitary symmetry of an isotropic oscillator \cite{fh}
(see also a closely related  description \cite{turbiner}).
The hidden symmetry
of the generalized  Calogero-Moser
system with a Coulomb potential is provided by a Dunkl deformed Runge-Lenz vector
\cite{CalCoul,Runge}.  Recently,  a Dunkl analog for the Dirac operator and its symmetries has
been investigated \cite{dirac}.

 At the equilibrium point corresponding to the potential minimum,  the dynamical
variables become frozen.
The Hamiltonian
reduces to the generalized Polychronakos-Frahm chain with the coordinate exchanges instead
of the spin ones.
The application of the exchange operators to discrete systems has a long story.
Using them, the commuting invariants for the   Haldane-Shastry  and Polychronakos-Frahm spin chains
have been constructed proving the integrability  \cite{minahan,poly93}.
The symmetry was expanded   to the (non-Abelian) Yangian  algebra \cite{yangian}, \cite{bernard}.
Its generators may be obtained from the symmetry algebra of the parental
Calogero-Sutherland and Calogero models.
Because of the  loss of continuous variables,  this map eliminates a lot of symmetries.
In particular, the quantum determinant of the Yangian algebra,
which combines the commuting integrals containing the Hamiltonian,
become a trivial constant in the freezing limit.
Instead, the power-series expansion in  Plank's constant was applied
to the quantum determinant. Then the first-order term provides the discrete
models with another set of Liouville integrals \cite{talstra,mathieu}.
This fact suggests that the conservation laws for the integrable chains with inverse-square interactions
have more elaborate structure than those for their dynamical counterparts and are needed for
further study.

In this article we describe the symmetry of the $N$-site generalized Polychronakos-Frahm chain
stressing out on its relation to the  $U(N)$ symmetry of the  isotropic oscillator, which is responsible
for the maximal superintegrability.
We start from the Dunkl-operator deformation of the unitary group generators, which provides
the dynamical and hidden symmetries for the generalized Calogero model.
At the lowest potential values, the dynamical degrees of freedom become limited to the discrete jumps
 between degenerate minima giving rise to a chain model.
We continue by extracting the conserving quantities for the discrete system
from the deformed unitary generators, wherein the diagonal elements correspond
to the  previously constructed integrals.
Algebraic relations between the reduced generators are derived also.
The symmetric combinations of the invariants  produce  integrals of the Polychronakos-Frahm chain.
We finish by inspecting the Coulomb confining potential in this context. The analog of the Runge-Lenz
vector vanishes at the discrete level and does not lead to a conservation law.

\section{Generalized Calogero model}
The Calogero model is an integrable  system of one-dimensional identical particles
 interacting by an inverse-square potential and bound by
an external harmonic force.  It is given by the Hamiltionain \cite{cal},
\begin{equation}
\label{H}
\hat {H}_\text{C}= \frac12\sum_{i=1}^N(\hat p_i^2+ x_i^2)+ \sum_{i<j}\frac{g(g\mp \hbar) }{ x_{ij}^2}.
\end{equation}
Here, $\hat p_i=-\imath\hbar\partial_i$ is the momentum operator,
$g$ is a coupling constant,
and the $-(+)$ sign in the potential corresponds to the bosons (fermions).
We use  a conventional notation for the particle distance, $ x_{ij}= x_i- x_j$.
The oscillator frequency is set to unity.

Most properties of the Calogero model (and its various  extensions), like
(super)integrability, spectrum, wave functions, and conservation laws are conditioned by its slightly modified version known as
a generalized Calogero  Hamiltonian \cite{poly92,brink},
\begin{equation}
\label{Hgen}
\hat H
= \frac12\sum_{i=1}^N\left( \hat p_i^2+  x_i^2\right)
+\sum_{i<j}\frac{g(g- \hbar M_{ij})}{ x_{ij}^2}.
\end{equation}
It includes a  permutation operator $M_{ij}$, which exchanges the coordinates
of $i$th and $j$th particles. At a first glance, such a modification may look
rather strange due to its nonlocality but for the identical particles it just reduces
to the standard Calogero Hamiltonian \eqref{H}. The advantage of the above
Hamiltonian is the representation in terms of the deformed momentum operator with a
derivative replaced  with the Dunkl operator,
\begin{gather}
\label{Hpi}
\hat H=\frac12\sum_{i=1}^N\left(\hat\pi_i^2+ x_i^2\right),
\\[2mm]
\label{pi}
\hat \pi_i=\hat p_i+\sum_{j\ne i}\frac{\imath g}{ x_{ij}} M_{ij}.
\end{gather}
The latter can be considered as a kind of flat nonlocal covariant derivative with the
following algebra \cite{dunkl}:
\begin{gather}
\label{com-pi}
[\hat\pi_i,\hat\pi_j]=[ x_i, x_j]=0,
\qquad
[ x_i,\hat\pi_j]= \imath \hat S_{ij},
\\[2mm]
\label{Sij}
\hat S_{ij}=(\delta_{ij}-1)gM_{ij}+\delta_{ij}\Bigg(\hbar+
  g\sum_{k\ne i} M_{ik}\Bigg).
\end{gather}
At the $g\to 0$ limit when the Calogero inverse-square potential is absent,
the operators $\hat\pi_i$,  $\hat S_{ij}$ are mapped to the
$\hat p_i$ and $\hbar\delta_{ij}$, correspondingly,  and the above relations
are reduced to the Heisenberg algebra commutation rules.

A Dunkl-operator analog of lowering-rising operators is defined in the standard way
\cite{brink,poly92,min-poly},
\begin{gather}
\label{hapm}
\hat a^\pm_i = \frac{ x_i\mp\imath\hat\pi_i }{\sqrt{2}},
\\
\label{com-ha}
[\hat a_i,\hat a_j]=[\hat a_i^+,\hat a_j^+]=0,
\qquad
[\hat a_i,\hat a_j^+]=\hat S_{ij}.
\end{gather}
The generalized  Hamiltonian can be expressed in terms of them,
\begin{equation}
\label{Ha}
\hat H=\frac{ 1}{2}\sum_i(\hat a_i^+\hat a_i+\hat a_i\hat a_i^+)
=\sum_i \hat a_i^+\hat a_i +\frac{\hbar N}{2}-S.
\end{equation}
Here,   by $S$, the rescaled  invariant of the permutation group algebra is denoted,
\be
\label{S}
S=-g\sum_{i<j}M_{ij},
\qquad
[S,M_{ij}]=0.
\ee
As a result, the lowering-rising operators obey a standard spectrum generating
relations \cite{brink,min-poly},
\be
\label{comHha}
[\hat H,\hat a_i^\pm]=\pm\hbar \hat a_i^\pm.
\ee
So, any bilinear combination,
\begin{equation}
\label{Eij}
\hat E_{ij}=\hat a_i^+\hat a_j,
\end{equation}
satisfies the conservation law,
\be
\label{comHE}
[\hat H,\hat E_{ij}]=0.
\ee

The elements $\hat E_{ij}$  are Dunkl-operator deformations of the unitary group generators.
Together with permutations $M_{ij}$, they provide
entire algebra of symmetries for the generalized Calogero model.
In addition, the following quadratic relation takes place among
$\hat E_{ij}$ and $\hat S_{ij}$ \cite{fh}:
\be
\label{cros}
\hat E_{ij} (\hat E_{kl}+\hat S_{kl})= \hat E_{il} (\hat E_{kj}+\hat S_{kj}).
\ee
The latter implies, in particular, the following commutation relation:
\begin{equation}
\label{comEij}
[\hat E_{ij}, \hat E_{kl}+\hat S_{kl}]= \hat E_{il}\hat S_{kj} - \hat S_{il} \hat E_{kj} .
\end{equation}

As a consequence, the diagonal elements are closed under commutation. Unlike
the  Cartan algebra, they are not Abelian but obey a simple commutation,
\begin{equation}
\label{comEi}
[\hat E_{ii}, \hat E_{kk}]=(\hat E_{ii}-\hat E_{kk})\hat S_{ik}.
\end{equation}
 The above algebra
ensures that the power sums  form a system of Liouville integrals of the Calogero system
\cite{poly92},
\be
\label{hcalE}
\hat{\cal E}_k=\sum_i\hat E_{ii}^k,
\qquad
[\hat {\cal E}_k,\hat {\cal E}_l]=0.
\ee
The generalized Hamiltonian itself is expressed in terms of the first member in this family,
\be
\label{HcalE}
\hat H=\hat{\cal E}_1-S+\frac{N\hbar}{2}.
\ee
Moreover,  it is a unique Casimir element (up to a nonessential constant term)
of the Dunkl-deformed $u(N)$ algebra \cite{fh}.

Remember that the same algebra \eqref{comEi} describes also the symmetries of the generalized
Hamiltonian related to the Sutherland model \cite{poly92}, an analog of the Calogero-Moser system with
trigonometric interactions \cite{suther}.

The antisymmetric  combinations of $\hat E_{ij}$ yield the
Dunkl  angular momentum components \cite{feigin,kuznetsov},
\be
\label{Lij}
\hat L_{ij}=\hat E_{ij}-\hat E_{ji}= x_i\hat\pi_j- x_j\hat\pi_i.
\ee
Together with permutations, they produce a deformation of $so(N)$ algebra
with a unique  Casimir element given by the  Dunkl angular momentum square
${\hat L}^2=\sum_{i<j}\hat L_{ij}^2$ shifted by a permutation invariant term
  \cite{fh},
\be
\label{hLsq}
\hat {\cal L}_2=\hat L^2+S^2-\hbar (N-2)S,
\qquad
[\hat L_{ij},\hat{\cal L}_2]=0.
\ee
It can be considered as a generalized angular Calogero Hamiltonian, which
 is reduced to the angular part of the Calogero model for identical particles \cite{flp},
\be
\begin{aligned}
\label{angular}
\hat H=-\frac{\hbar^2}{2}\left( \partial_{ r}^2 + \frac{N-1}{ r}\partial_{ r}\right)
+\frac{ r^2}{2}+\frac{\hat {\cal L}_2}{2 r^2}
\\
\qquad
 \text{with}
 \quad
  r=\sqrt{ x^2}.
  \end{aligned}
\ee

The symmetric combinations of the deformed $u(N)$ generators produce
a Dunkl-operator deformation for the well-known Fradkin tensor \cite{francisco},
\be
\label{Iij}
\hat I_{ij}=\hat E_{ij}+\hat E_{ji}-\hat S_{ij}=\hat\pi_i\hat\pi_j+ x_i x_j.
\ee
Remember that the angular momentum and Fradkin tensor
describe, respectively, the dynamical and hidden symmetries
of the $N$-dimensional isotropic oscillator \cite{fradkin}.

The diagonal algebra \eqref{comEi}  has an Abelian basis obtained by applying a shift  to its elements.
The shift  is a  tail composed of  exchange operators  \cite{bernard},
\begin{equation}
\label{Di}
\hat D_i=\hat E_{ii}-\hat S_i, 
\qquad
\hat S_i=\sum_{j=1}^{i-1}\hat S_{ij},
\end{equation}
where $\hat S_1=0$ is supposed.
Together with permutations, the elements $\hat D_i$ satisfy the defining relations of
 degenerate affine Hecke algebra,
\be
\label{Hecke}
\begin{gathered}
{[}\hat D_i,\hat D_j]=0 ,
\qquad
[\hat D_k,\hat S_{jj+1}] =0 \quad \text{if\; $k\ne i,i+1$},
\\
\hat D_{i+1} \hat S_{jj+1}-\hat S_{jj+1}\hat D_i = g^2.
\end{gathered}
\ee
Note that the tails $S_i$  satisfy the same relations; i.e.,  the above equations remain true upon the substitution
$\hat D_i\to \hat S_i$.
As a result, the modified diagonal elements  can be considered as an analog of
Liouville integrals for the generalized Hamiltonian \eqref{Hgen}, which may be expressed via them
 using the representation \eqref{Ha},
\be
\label{HD}
\hat H=\sum_i \hat D_i +\frac{\hbar N}{2}.
\ee
The higher-order power sums  define the higher Hamiltonians,
\be
\label{calD}
\hat{\cal D}_k=\sum_{i}\hat D_i^k.
\ee
The second member, $\hat{\cal D}_2$, corresponds to the generalized Calogero-Sutherland model.
Note that in contrast to the previous integrals \eqref{hcalE}, the permutation invariance
 is not evident but can be verified.
More familiar are the monomials  given by the generating function $\prod_i(u-\hat D_i)$
\cite{bernard}.
In general, any symmetric polynomial in $\hat{\cal D}_i$ is permutation invariant and
reduced to the constant of motion of the Calogero model \eqref{H} for indistinguishable
particles.

More  symmetric polynomial  may include the nondiagonal
components of the generators $\hat E_{ij}$.
The permutations can be used in such expressions as well.
Here  are some simplest examples of such integrals,
\be
\label{hIsq}
\sum_{i, j} \hat I_{ij}^k,
\quad
\sum_{i, j} \hat L_{ij}^{2k},
\quad
\sum_{i<j} \hat I_{ij}^kM_{ij},
\quad
\sum_{i,j} \hat I_{ii}^k\hat L_{ij}^{2l}.
\ee

\section{Generalized Polychronakos-Frahm chain}
Let us set the interaction constant to \emph{unity}, $g=1$. In the current section, we
consider the generalized Calogero model at the \emph{equilibrium} point, where
the (classical) confining      Calogero potential takes its minimal value,
 \be
\label{min}
 \qquad
\frac{\partial V}{\partial{x_i}}=0,
 \qquad
 V(x)=\sum_{i=1}^N \frac{x_i^2}{2}+\sum_{i<j}\frac{1}{x_{ij}^2}.
\ee
The coordinates are given by the roots of the $N$th-order
Hermite polynomial \cite{frahm}.
All roots differ, so there are $N!$
equivalent minima connecting by the coordinate exchanges $M_{ij}$.
These are only allowed evolutions in the frozen system.

Consider the expansion of the  Hamiltonian \eqref{Hgen}
in powers of Planck’s constant,
\be
\label{hH}
\hat H
=V+\hbar  H^{(1)}-\frac{ \hbar^2 \partial^2}{2}.
\ee
The first-order term can be considered as a generalized
Polychronakos-Frahm Hamiltonian, where the spin exchange operator is replaced
by the coordinate exchange \cite{poly93},
\be
\label{H1}
 H^{(1)}=\sum_{i<j}\frac{1}{x_{ij}^2}M_{ij}.
\ee
The $SU(n)$ symmetric  spin chain is recovered from the above Hamiltonian after
the replacement of the coordinate permutations with spin exchange operators
so that
\be
\label{PF}
 H_\text{PF}=\sum_{i<j}\frac{P_{ij}}{x_{ij}^2}.
\ee
Here, $P_{ij}$ permutes the $i$th and $j$th spins, which take values in the fundamental
representation of the $SU(n)$ group.
Both Hamiltonians become identical on the bosonic (fermionic) states provided that
the particles are endowed with additional spin  degrees of
freedom. Then the entire wave function must be symmetric (antisymmetric)
under simultaneous exchanges of coordinates and spins for bosons (fermions).
 A permutation of spatial coordinates  $M_{ij}$
can be replaced by the spin exchange operators
$P_{ij}$ and $-P_{ij}$  in the bosonic and fermionic cases respectively.
Note that the projection inverts the order of permutations so that the operator $M_{ij}M_{kl}$
must be substituted by $P_{kl}P_{ij}$.

Let us now construct the constants of motion of the  chain Hamiltonian \eqref{H1} by applying
the $\hbar$ expansion  to the integrals of the original dynamical model \eqref{Hgen} at the equilibrium
point.  They do not provide directly the invariants of the  $SU(n)$ spin chain \eqref{PF}.
In order to get them, one needs to carry out
symmetrization over all particles prior to the projection.
For instance, $H^{(1)}$ stays invariant under a selected coordinate
exchange,  $M_{ij}$, but $H_\text{PF}$ does not preserve its spin counterpart, $P_{ij}$. However, both Hamiltonians
 preserve the symmetrized
version given by the element $S$ \eqref{S}.

The Dunkl momentum \eqref{pi} and  permutation matrix \eqref{Sij} can be presented as follows:
\begin{align}
\hat\pi_i&=\pi_i-\imath\hbar \partial_i
\quad
\text{with}
\quad
\pi_i=\sum_{j\ne i}\frac{\imath}{x_{ij}} M_{ij},
\\
\hat S_{ij}&=S_{ij}+\hbar\delta_{ij}
\quad
\text{with}
\\[3pt]
&\qquad\qquad S_{ij}=(\delta_{ij}-1)M_{ij}+\delta_{ij} \sum_{k\ne i} M_{ik}.
\nonumber
\end{align}
Here and in the following, the superscript is omitted
in the zeroth-order term of any operator, so that $\pi_i^0=\pi_i$.
In that limit, the canonical commutation relations resemble their original form
\eqref{com-pi},
\be
\label{com-pi0}
\begin{gathered}
{[}\pi_i ,\pi_j]=0,
\qquad
[x_i,\pi_j]= \imath S_{ij},
\\[4pt]
S_{ij}=
\begin{cases}
  -M_{ij}, & \mbox{if } i\ne j, \\
  \sum_{k\ne i} M_{ik}, & \mbox{otherwise}.
\end{cases}
\end{gathered}
\ee

Recall now that the particle coordinates are set by the roots of the Hermite polynomial,
which imposes certain algebraic relations on them (see, for example, \cite{calo,mathieu}).
As a result, the discrete Dunkl momenta are not independent any more but undergo
 additional algebraic constraints. In particular, the following relations hold among the fixed
 phase space variables:
\begin{gather}
\label{sum-x}
\sum_i x_i=\sum_i \pi_i=\sum_i S_{ik}=0,
\\
\label{xp}
\begin{gathered}
x^2=\pi^2=\sum_{i<j}\frac{2}{x_{ij}^2}=\sfrac12N(N-1),
\\
x\cdot \pi =- \pi\cdot x=-\imath S.
\end{gathered}
\end{gather}

 In general, all relations between the operators of the dynamical Calogero system are preserved
at the $\hbar=0$ limit.
%
In particular, the frozen  lowering-rising
operators
\be
\label{apm}
a^{\pm}_i = \frac{x_i \mp \imath \pi_i}{\sqrt 2},
\qquad
\hat a^{\pm}_i =a_i^\pm\pm\frac{\hbar}{\sqrt{2}}\partial_i
\ee
obey a  rule similar to the commutations of the deformed Heisenberg algebra \eqref{com-ha}
 \cite{poly93},
\be
\label{com-a}
[ a_i,a_j]=[ a_i^+,a_j^+]=0,
\qquad
[ a_i,a_j^+]=S_{ij}.
\ee
Because of the minimum condition \eqref{min},  the spectrum generating relation \eqref{comHha}
remains valid for the generalized Polychronakos-Frahm
chain too \cite{poly93},
\be
\label{comHa}
[H^{(1)}, a_i^\pm]=\pm  a_i^\pm.
\ee
However,   unlike the dynamical case, the discrete Hamiltonian
is not expressed via lowering-rising operators [see Eq.\eqref{Ha}].

The  constants of motion of the generalized Calogero model \eqref{Eij} have terms up
to the second order in their expansion,
\be
\label{hEij}
\hat E_{ij}=E_{ij}+\hbar  E_{ij}^{(1)}-\frac{\hbar^2}{2} \partial_i\partial_j.
\ee

The relations \eqref{comHa} imply conservation of the   constant terms,
\be
\label{E0}
E_{ij}=a_i^+a_j,
\qquad
[H^{(1)},E_{ij}]=0.
\ee

For the Dunkl angular momentum \eqref{Lij}, the  $\hbar^2$ part
vanishes,
while the $\hbar$ term corresponds to the usual angular
momentum operator in quantum mechanics,
\be
\label{expLij}
\hat L_{ij}= L_{ij}- \imath\hbar (x_i\partial_j-x_j\partial_i),
\qquad
L_{ij}=x_i\pi_j-x_j\pi_i.
\ee
A similar expansion   for the Fradkin tensor is more complex,
\be
\label{expIij}
\hat I_{ij}=I_{ij}+\hbar I_{ij}^{(1)}-\hbar^2\partial_i\partial_j,
\qquad
I_{ij}=x_ix_j+\pi_i\pi_j.
\ee
The first order operator-valued coefficient is given by
\begin{align}
\label{I1ij}
&I_{ij}^{(1)}
=
\frac{1}{x_{ij}^2}M_{ij}
+\sum_{k\ne i,j }
\left(
\frac{\partial_i}{x_{jk}}M_{jk}+\frac{\partial_j}{x_{ik}}M_{ik}
\right)
\end{align}
for $ i\ne j$ and
\begin{align}
\label{I1ii}
&I_{ii}^{(1)}=
\sum_{k\ne i}\frac{1}{x_{ik}}\left(
\partial_i+\partial_k-\frac{1}{x_{ik}}
\right)M_{ik}.
\end{align}

As was discussed above, the algebraic relations between the symmetry generators
of the dynamical system remain true at the freezing limit. In particular,
the  most general relation \eqref{cros} and its consequences \eqref{comEij}, \eqref{comEi}
are reduced, respectively, to the following equations:
\begin{gather}
\label{cros0}
E_{ij} (E_{kl}+S_{kl})= E_{il} (E_{kj}+S_{kj}),
\\[2mm]
\label{comE0}
[E_{ij}, E_{kl}+S_{kl}]= E_{il}S_{kj} - S_{il} E_{kj} ,
\\[2mm]
\label{comEi0}
[E_{i}, E_{k}]=(E_{i}-E_{k})S_{ik}.
\end{gather}
The power sums of diagonal elements  yield  Liouville integrals of the
 Polychronakos-Frahm chain \cite{poly93},
\be
\label{calE}
{\cal E}_k=\sum_i E_{ii}^k,
\qquad
[ {\cal E}_k, {\cal E}_l]=0,
\qquad
[H^{(1)},{\cal E}_k]=0.
\ee
The first element of this set is rather trivial, ${\cal E}_1=S+\frac{N(N-1)}{2}$, as is easy to get using the equations
\eqref{xp} and \eqref{apm}. The higher rank ${\cal E}_k$ have more complicated expressions.

For the dynamical system, the quadratic relations  \eqref{cros} are the
only constraints which the symmetry generators $\hat E_{ij}$ obey \cite{fh}.
However, there are a lot of other restrictions on them at the equilibrium point.
For example, the Eqs. \eqref{sum-x} imply the sum vanishing rules,
\be
 \sum_{i}E_{ik}=\sum_{i}E_{ki}= \sum_{i}L_{ik}= \sum_{i}I_{ik}=0.
\ee

In the dynamical case, the angular Calogero Hamiltonian \eqref{hLsq} plays an important role
among constants of motion.  In the absence of an oscillator potential,  it maps the Liouville
set to additional integrals.
However, in the equilibrium limit, the  angular part
does not produce a new integral but just is expressed via trivial ones.
Using the relations \eqref{sum-x}, \eqref{xp}, \eqref{Lij}, and \eqref{Lsq}, it is easy to verify that
the operator $\hat {\cal L}_2$ is a scalar at the equilibrium. Its $\hbar$-linear
coefficient reproduces the chain Hamiltonian as was argued earlier \cite{flp},
\be
\label{Lsq}
{\cal L}_2
=r^4,
\qquad
{\cal L}_2^{(1)}=2r^2H^{(1)}
\qquad
\ee
with  $r^2=\frac12N(N-1)$  \eqref{xp}.

Let us calculate also a chain analog of the Fradkin's tensor square. It corresponds to
the second ($k=2$) member of the first sequence  presented in \eqref{hIsq},
\be
\sum_{i,j}I_{ij}^2
=\sum_{i\ne j\ne k}M_{ijk}-2S^2
+\sfrac12N(N-1)(N^2-N+4).
\ee
Here, $M_{ijk}$ is a cyclic permutation of the marked three coordinates.
It also is expressed through the invariants of the permutation group algebra.
Nevertheless, we expect that higher degree power sums
from \eqref{hIsq}  at the equilibrium give rise  to nontrivial integrals of motion
for the chain Hamiltonians \eqref{H1}, \eqref{PF}.

Finally, consider  the shifted diagonal elements \eqref{Di}. At the freezing level,
they also commute,
\begin{gather}
\label{comD0}
{[}D_i,D_j]=0,
\qquad
D_i=E_{ii}-S_i.
\end{gather}
Together with permutations, they form also  degenerate affine Hecke algebra \eqref{Hecke} (with $g=1$).
Contrary to the nonshifted case \eqref{E0}, \eqref{calE},  the related
symmetric polynomials, $\mathcal{D}_k= \sum_i D_i^k$, are scalars (multiples of identity) and do not lead to
a  conservation law.
At the same time, the $\hbar$-linear terms of the dynamical integrals \eqref{calD}
form a family of commuting nontrivial integrals  \eqref{calE},
\begin{gather}
\mathcal{D}_k^{(1)}=\sum_{i}\sum_{l=0}^{k-1}  D_i^l D^{(1)}_i D_i^{k-l-1},
\quad
[ \mathcal{D}_k^{(1)}, \mathcal{D}_l^{(1)}]=0,
\end{gather}
where $D^{(1)}_i=\frac12 I^{(1)}_{ii}$.
The first element of the family describes the chain Hamiltonian: $\mathcal{D}^{(1)}_1=-2H^{(1)} $.
This remarkable property was established first  for the Haldane-Shastry chain  using the Yangian represtation
\cite{talstra} and has been  extended later to the Polychonakos-Frahm chain
\cite{mathieu}.

\section{Relation to generalized Calogero-Coulomb system}
One can  start from another radial confining potential in the dynamical  system \eqref{Hgen}
 in order to get  its discrete analog at the semiclassical limit \eqref{H1}.
 Apart from the oscillator case,  the Coulomb potential is  preferable among others
 since it leads to a superintegrable system as well \cite{CalCoul},
\be
\label{Hcoul}
\hat H_\text{Coul}=\sum_{i=1}^N\frac{\hat\pi_i^2}{2}-\frac{1}{ r}.
\ee
So, there is a hope to extract
 more integrals for the chain system as a reduction of the recently obtained
 Dunkl-operator analog of
the Runge-Lenz vector \cite{Runge},
\be
\label{hA}
\hat A_i=\frac12\sum_j\{\hat L_{ij},\hat \pi_j\}+\frac{\imath}{2}[\hat \pi_i,\hat S]-\frac{ x_i}{r}.
\ee
Together with the Dunkl angular momentum $\hat L_{ij}$, it yields the Dunkl-operator
deformation of the $SO(N+1)$ group generators, which describe the
symmetry of the generalized Calogero system with Coulomb potential \eqref{Hcoul}
\cite{Runge}.

The equilibrium point is defined now by  a solution of the  equations,
 \be
\label{min-coul}
 \frac{\partial V_\text{Coul}}{\partial{y_i}}=0,
 \;\;\;
 V_\text{Coul}(y)= -\frac{1}{r_y}+\sum_{i<j}\frac{1}{y_{ij}^2},
 \;\;\;
 r_y^2=x^2.
\ee
In order to separate it from the oscillator minimum \eqref{min}, the Coulomb point is marked by
$y$. It can be obtained from the oscillator point, given by zeros of Hermite polynomial \eqref{min},
by
the  map $y_i= r^3 x_i$. The latter implies that $r_y=r^4$. Therefore, passing from the oscillator
to Coulomb potential, the chain Hamiltonian
\eqref{H1} just undergoes a rescaling
$
H^{(1)}_\text{Coul}=r^{-6}H^{(1)}
$.
Evidently,  the Dunkl momentum transforms as  $ r^{-3} \pi_i\to \pi_i$. The relations \eqref{sum-x}, \eqref{xp}
stay unchanged apart from the first equation in \eqref{xp}, which must be  replaced now by
\be
\label{yp}
\frac{1}{r_y}=\pi^2=\sum_{i<j}\frac{2}{y_{ij}^2}.
\ee
Using them, it is easy to verify that the Dunkl deformed
Runge-Lenz vector vanishes trivially at the Coulomb equilibrium point,
\be
\label{A}
A_i=\frac12\sum_j\{L_{ij},\pi_j\}+\frac{\imath}{2}[\pi_i,S]-\frac{y_i}{r_y}=0.
\ee
This fact eliminates the power sums $\hat {\cal A}_k=\sum_i \hat A_i^k$
at the equilibrium,  $ {\cal A}_k=0$, but does not necessarily imply the
vanishing of the first-order operator in powers of $\hbar$. Actually, the first two
such operators are
\be
\label{A1}
{\cal A}^{(1)}_1=0,
\qquad
{\cal A}^{(1)}_2=S/r_y,
\ee
as can be derived from the definition \eqref{hA}.
Note that the square of the Dunkl Runge-Lenz vector
may be shifted certain way to become compatible
with the Dunkl angular momentum.
Moreover, it provides a deformation of the well-known relation
between the  angular momentum and Runge-Lenz squares  \cite{Runge},
\begin{gather}
\label{Asq}
 \hat {\cal A}'_2=\hat {\cal A}_2+2\hbar S \hat H_\text{Coul},
 \qquad
 [\hat L_{ij},\hat {\cal A}_2]=0,
 \\
 \label{hAsq}
\hat {\cal A}'_2=2\hat H_\text{Coul}\hat  {\cal L}_2
+\frac{\hbar^2(N-1)^2}{2} \hat H_\text{Coul}+1.
 \end{gather}
 %
Applying  the Taylor expansion with the coefficients written in terms of
$y$ coordinates,
\be
{\cal L}_2=r_y,
\quad
{\cal L}_2^{(1)}=2r_y^2 H_\text{Coul}^{(1)},
\quad
V_\text{Coul}=-\frac{1}{2r_y},
\ee
 we obtain the vanishing condition, $ {\cal A}_2'^{(1)}=0$, which  is equivalent to  the second equation in \eqref{A1}.

\acknowledgments
The author is grateful to Armen Nersessian for stimulating discussions.
This work was   supported by the Armenian
State Committee of Science Grants No. SFU-02, No. 18RF-002, and No. 18T-1C106.
It was fulfilled within the ICTP Affiliated Center Program AF-04.


\begin{thebibliography}{99}

\bibitem{haldane88}
F.D.M. Haldane,
\emph{Exact Jastrow-Gutzwiller resonating-valence-bond ground state of the
spin-1/2 antiferromagnetic Heisenberg chain with $1/r^2$ exchange},
 \href{https://doi.org/10.1103/PhysRevLett.60.635}{Phys. Rev. Lett. {\bf60} (1988) 635}.

\bibitem{shastry88}
B.S. Shastry,
\emph{Exact solution of an $S=1/2$ Heisenberg antiferromagnetic chain with long-ranged interactions},
\href{https://doi.org/10.1103/PhysRevLett.60.639}{Phys. Rev. Lett.  {\bf 60} (1988) 639}.

\bibitem{suther}
B. Sutherland,
\emph{Exact Results for a Quantum Many-Body Problem in One Dimension},
\href{https://doi.org/10.1103/PhysRevA.4.2019}{Phys. Rev. A {\bf 4}  (1971) 2019};
\emph{Exact Results for a Quantum Many-Body Problem in One Dimension. II},
\href{https://doi.org/10.1103/PhysRevA.5.1372}{A {\bf 5}  (1972) 1372}.


\bibitem{cal}
F.~Calogero,
\emph{Solution of a three-body problem in one dimension},
\href{http://dx.doi.org/10.1063/1.1664820}{J. Math. Phys. {\bf 10} (1969) 2191};
\emph{Solution of the one-dimensional $N$-body problems with quadratic and/or inversely quadratic pair potentials},
 \href{http://dx.doi.org/10.1063/1.1665604}{{\sl ibid.} {\bf 12} (1971) 419}.

\bibitem{poly93}
A.~Polychronakos,
\emph{Lattice Integrable Systems of Haldane-Shastry Type},
Phys. Rev. Lett. {\bf 70} (1993) 2329,
\href{http://arxiv.org/abs/hep-th/9210109}{hep-th/9210109}.

\bibitem{frahm}
H. Frahm,
\emph{Spectrum of a spin chain with inverse square exchange},
 J. Phys. A {\bf 26} (1993) L473,
\href{https://arxiv.org/abs/cond-mat/9303050}{cond-mat/9303050}.

\bibitem{min-poly}
J.A. Minahan and A.P. Polychronakos,
\emph{Integrable Systems for Particles with Internal Degrees of Freedom},
Phys. Lett. B {\bf 302} (1993) 265,
\href{https://arxiv.org/abs/hep-th/9206046}{hep-th/9206046}.

\bibitem{khare}
A.~Khare,
\emph{Exact solution of an $N$-body problem in one dimension},
J. Phys. A {\bf 29} (1996) L45,
\href{http://arxiv.org/abs/hep-th/9510096}{hep-th/9510096}.

\bibitem{rev-olsh}
M.~A.~Olshanetsky and  A.~M.~Perelomov,
\emph{Classical integrable finite dimensional systems related to Lie algebras},
\href{http://dx.doi.org/10.1016/0370-1573(81)90023-5}{Phys. Rept.  {\bf 71} (1981) 313};
\emph{Quantum integrable systems related to Lie algebras},
\href{http://dx.doi.org/10.1016/0370-1573(83)90018-2 }{{\sl ibid.}  {\bf 94} (1983) 313}.

\bibitem{rev-poly}
A.~P.~Polychronakos,
\emph{Physics and mathematics of Calogero particles},
J. Phys. A  {\bf 39}  (2006) 12793,
\href{http://arxiv.org/abs/hep-th/0607033}{hep-th/0607033}.

\bibitem{moser}
J. Moser,
\emph{Three integrable Hamiltonian systems connected with isospectral deformations},
\href{http://dx.doi.org/10.1016/0001-8708(75)90151-6}{Adv. Math. {\bf 16} (1975) 197}.


\bibitem{woj83}
 S.~Wojciechowski,
 \emph{Superintegrability of the Calogero-Moser system},
 \href{http://dx.doi.org/10.1016/0375-9601(83)90018-X}{Phys. Lett. A {\bf 95} (1983) 279}.

\bibitem{kuznetsov}
V.B.~Kuznetsov,
\emph{Hidden symmetry of the quantum Calogero-Moser system},
Phys. Lett. A {\bf 218} (1996) 212,
\href{http://arxiv.org/abs/solv-int/9509001}{solv-int/9509001}.

  \bibitem{gonera}
  C. Gonera,
 \emph{A note on superintegrability of the quantum Calogero model},
\href{https://doi.org/10.1016/S0375-9601(98)00903-7}{Phys. Lett. A {\bf237} ( 1998) 365}.

  \bibitem{kosinski}
  C. Gonera and P. Kosinski,
  \emph{Calogero model and sl(2,R) algebra},
  Acta Phys. Polon. B {\bf 30} (1999) 907,
 \href{http://arxiv.org/abs/hep-th/9810255}{hep-th/9810255}.

\bibitem{CalCoul}
  T.~Hakobyan, O.~Lechtenfeld, and A.~Nersessian,
\emph{Superintegrability of generalized Calogero models with oscillator or Coulomb potential},
   Phys. Rev. D \textbf{90}  (2014) 101701(R),
 \href{http://arxiv.org/abs/1409.8288}{arXiv:1409.8288}.

\bibitem{dunkl}
C.~F.~Dunkl,
\emph{Differential-difference operators associated to reflection groups},
\href{http://www.ams.org/journals/tran/1989-311-01/S0002-9947-1989-0951883-8}%
{Trans. Amer. Math. Soc. {\bf 311} (1989) 167}.

\bibitem{poly92}
A.~Polychronakos,
\emph{Exchange operator formalism for integrable systems of particles},
Phys. Rev. Lett. {\bf 69} (1992) 703,
\href{http://arxiv.org/abs/hep-th/9202057}{hep-th/9202057}.

\bibitem{brink}
L.~Brink, T.~Hansson, and M.~Vasiliev,
\emph{Explicit solution to the N-body Calogero problem},
Phys. Lett. B {\bf  286} (1992) 109,
\href{http://arxiv.org/abs/hep-th/9206049}{hep-th/9206049}.

\bibitem{fh}
 M.~Feigin and T.~Hakobyan,
\emph{On the algebra of Dunkl angular momentum operators},
JHEP {\bf 11} (2015) 107,
\href{http://arxiv.org/abs/1409.2480}{arXiv:1409.2480}.

\bibitem{turbiner}
A. Turbiner,
\emph{Hidden algebra of the $N$-body Calogero problem},
Phys. Lett.  B {\bf320}  (1994) 281,
\href{https://arxiv.org/abs/hep-th/9310125}{hep-th/9310125}.

\bibitem{Runge}
T.~Hakobyan and A.~Nersessian,
\emph{Runge-Lenz vector in Calogero-Coulomb problem},
 Phys. Rev. A \textbf{92} (2015) 022111,
\href{http://arxiv.org/abs/1504.00760}{arXiv:1504.00760}.

\bibitem{dirac}
H. De Bie, R. Oste, and J. Van der Jeugt,
\emph{On the algebra of symmetries of Laplace and Dirac operators},
Lett. Math. Phys. {\bf 108} (2018) 1905,
\href{https://arxiv.org/abs/1701.05760}{arXiv:1701.05760};
\emph{The total angular momentum algebra related to the $S_3$ Dunkl Dirac equation},
Ann. Phys. {\bf 389} (2018) 192,
\href{https://arxiv.org/abs/1705.08751}{arXiv:1705.08751}.



\bibitem{minahan}
M. Fowler and J.A. Minahan,
\emph{Invariants of the Haldane-Shastry $SU(N)$ Chain},
Phys. Rev. Lett. {\bf 70} (1993) 2325,
\href{https://arxiv.org/abs/cond-mat/9208016}{cond-mat/9208016}.

\bibitem{yangian}
F.D.M. Haldane, Z.N.C. Ha, J.C. Talstra, D. Bernard and V. Pasquier,
\emph{Yangian symmetry of integrable quantum chains with long-range interactions and a new description of states in conformal field theory},
 \href{https://doi.org/10.1103/PhysRevLett.69.2021}{Phys. Rev. Lett. {\bf 69} (1992) 2021}.

\bibitem{bernard}
D. Bernard, M. Gaudin, F.D.M. Haldane, and V. Pasquier,
\emph{Yang-Baxter equation in spin chains with long range interactions},
J. Phys.  A {\bf 26} (1993) 5219,
\href{https://arxiv.org/abs/hep-th/9301084}{hep-th/9301084}.

\bibitem{mathieu}
P. Mathieu and Y. Xudous,
\emph{Conserved charges of non-Yangian type
for the Frahm-Polychronakos spin chain},
J. Phys. A {\bf34} (2001) 4197,
\href{https://arxiv.org/abs/hep-th/0008036}{hep-th/0008036}.

\bibitem{talstra}
J.C. Talstra and F.D.M. Haldane,
\emph{Integrals of motion of the Haldane Shastry Model},
J. Phys.  A {\bf 28} (1995) 2369,
\href{https://arxiv.org/abs/cond-mat/9411065}{cond-mat/9411065}.

\bibitem{feigin}
M. Feigin,
\emph{Intertwining relations for the spherical parts of generalized calogero operators},
\href{http://dx.doi.org/10.1023/A:1023231402145}{Theor. Math. Phys. {\bf 135} (2003) 497}.

\bibitem{flp}
M.~Feigin, O.~Lechtenfeld, and  A.~Polychronakos,
\emph{The quantum angular Calogero--Moser model},
JHEP {\bf 1307} (2013) 162,
\href{http://arxiv.org/abs/1305.5841}{arXiv:1305.5841}.

\bibitem{francisco}
F.~Correa, T.~Hakobyan, O.~Lechtenfeld, and A.~Nersessian,
\emph{Spherical Calogero model with oscillator/Coulomb potential: quantum case},
Phys. Rev. D {\bf 93} (2016)  125009,
\href{http://arxiv.org/abs/1604.00027}{arXiv:1604.00027}.

\bibitem{fradkin}
D.M. Fradkin,
\emph{Existence of the dynamic symmetries $O_4$ and $SU_3$ for all classical central potential problems},
\href{http://dx.doi.org/10.1143/PTP.37.798}{Theor. Phys. {\bf 37} (1967) 798};
\emph{Three-dimensional isotropic harmonic oscillator and $SU_3$},
\href{http://dx.doi.org/10.1119/1.1971373}{Am. J. Phys. {\bf 33} (1965) 207}.

\bibitem{calo}
F. Calogero,
\emph{Equilibrium Configuration of the One-Dimensional $n$-Body Problem
with Quadratic and Inversely Quadratic Pair Potentials},
\href{https://dx.doi.org/10.1007/BF02785163}{Lett. Nuvo Chim. {\bf 20} (1977) 251}.

\end{thebibliography}
 \end{document}